\documentclass[12pt]{article} % For LaTeX2e
\usepackage{times}

\usepackage{natbib, times}

\usepackage{amsmath,amsfonts,bm}

\def\eqref#1{equation~\ref{#1}}
\def\1{\bm{1}}

\DeclareMathAlphabet{\mathsfit}{\encodingdefault}{\sfdefault}{m}{sl}
\SetMathAlphabet{\mathsfit}{bold}{\encodingdefault}{\sfdefault}{bx}{n}

\usepackage{hyperref}
\usepackage{url}
\usepackage{subfig}
\usepackage{setspace}
\usepackage{authblk}

\author[1]{Stefan Vamosi}
\author[2]{Michael Platzer}
\author[1]{Thomas Reutterer}
\affil[1]{Vienna University of Economics and Business, Wirtschaftsuniversitat Wien, Vienna, Austria}
\affil[2]{MOSTLY AI Solutions, Hegelgasse 21/3, A-1010 Vienna, Austria}
\affil[ ]{stefan.vamosi@wu.ac.at, michael.platzer@mostly.ai, thomas.reutterer@wu.ac.at}

\title{AI-based Re-identification of Behavioral Clickstream Data}

\date{}

\usepackage{graphicx}
\graphicspath{ {.} }

\begin{document}

\maketitle

\begin{abstract}
AI-based face recognition, i.e., the re-identification of individuals within images, is an already well established technology for video surveillance, for user authentication, for tagging photos of friends, etc. This paper demonstrates that similar techniques can be applied to successfully re-identify individuals purely based on their behavioral patterns. In contrast to de-anonymi-\\zation attacks based on record linkage, these methods do not require any overlap in data points between a released dataset and an identified auxiliary dataset. The mere resemblance of behavioral patterns between records is sufficient to correctly attribute behavioral data to identified individuals. Further, we can demonstrate that data perturbation does not provide protection, unless a significant share of data utility is being destroyed. These findings call for sincere cautions when sharing actual behavioral data with third parties, as modern-day privacy regulations, like the GDPR, define their scope based on the ability to re-identify. This has also strong implications for the Marketing domain, when dealing with potentially re-identify-able data sources like shopping behavior, clickstream data or cockies. We also demonstrate how synthetic data can offer a viable alternative, that is shown to be resilient against our introduced AI-based re-identification attacks.

\end{abstract}

\vspace{1cm}


\section{Introduction}\label{sec:intro}

The steady rise of digital native business formats witnesses the tremendous opportunities offered by an exploding amount and variety of individual-level, behavioral micro-data accruing in a broad range of industries. For example, firms like Amazon, Netflix or Meta track the behavior of their customers to derive personalized recommendations and targeted marketing actions. Other companies realize that sharing customer information with other parties (e.g., linked with “Internet of Things” elements, such as mobile tracking meters, medical or fitness devices, etc.) can create synergies for both sides. Likewise, the non-profit sector and research institutions increasingly rely on the availability or “share-ability” of publicly available or open behavioral data \citep{beaulieu2019privacy}.

However, all these benefits are in strong contrast to the legitimate desire of individuals to protect their privacy and to refrain from sharing their personal data \citep{wieringa2021data}. In the vein of the Facebook-Cambridge Analytica scandal, also firms have increasingly become sensitive to protect their customer data against re-identification attacks and their brands against a loss in customer trust \citep{schneider2017protecting, schneider2018flexible}. All these concerns lead to modern privacy regulations (in particular EU’s GDPR and California’s CCPA) which impose very strict standards for data anonymization. Both the GDPR and the CCPA do not specify any specific process for anonymization, but they demand the outcome to be irreversible to prevent re-identification of individuals by all the means reasonably likely to be used \citep{finck2020they}.

Against this background, the key challenge for many firms is to keep benefiting from data-driven marketing while maintaining the privacy of their customers’ data. Previous research already showed that conventional perturbation techniques (such as adding random noise, masking, or obfuscation) fail to do so in the presence of high dimensional, highly correlated data, which typically arise when observing individuals over an extended period of time, i.e. for sequential personal data. For example, \cite{narayanan2006break} document successful re-identification attacks in the context of Netflix user’s movie ratings, \cite{de2013unique} for human mobility traces and\cite{de2015unique} for credit card retail transaction data.

In this research, we extend this perspective and show that a powerful general-purpose, AI-based model recently proposed by \cite{vamosi2021} is capable to re-identify behavioral data in a highly effective way. As we will demonstrate in detail, this makes standard “anonymization” techniques inapt to protect individual-level sequential data against a new breed of AI-based pattern attacks. We also show that data synthetization can help to manage the trade-off between preserving the useful information in the original data, while reducing the risk of violating privacy.

\section{Behavioral Pattern Attacks}

In 2006 Netflix released an "anonymized" dataset to participants of a Machine Learning competition. The dataset consisted of 470,000 Netflix users, and their complete history of over 100 million movie ratings across 18,000 movies. Despite the fact, that only a subset of the actual customer base has been released, despite the omission of all customer-level attributes, and despite the injection of random noise to the data, researchers were quickly able to perform and publish a successful re-identification attack \citep{narayanan2006break}. They achieved so by fuzzy matching the released de-identified records with publicly available, yet identified records, that were obtained from a popular movie rating website, IMDB. For every successful linkage they would then have been able to expose the complete Netflix rating history of users, who only shared a fraction of their ratings publicly via IMDB.

\begin{figure}%
    \centering
    \subfloat[\centering Linkage attacks rely on an overlap of the data points of the released and the auxiliary data.]{{\includegraphics[width=5cm]{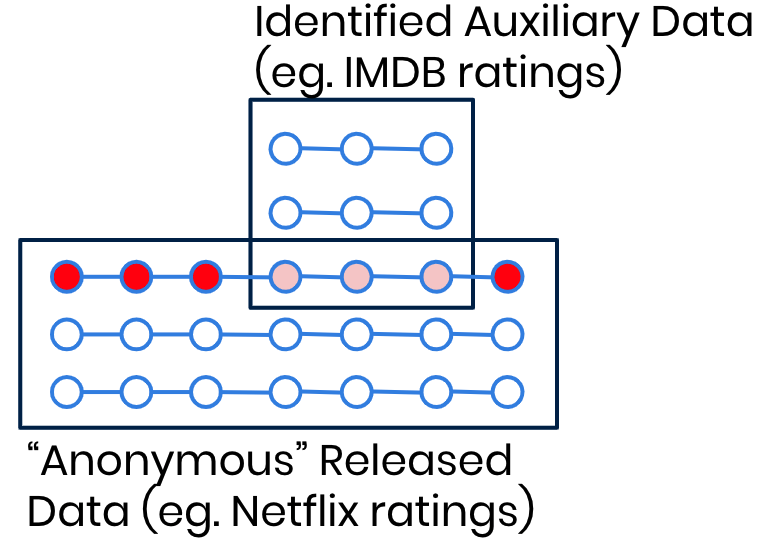} }}%
    \qquad
    \subfloat[\centering Pattern attacks do not require an overlap of the data points, but merely of the data subjects of the released and the auxiliary data.]{{\includegraphics[width=5cm]{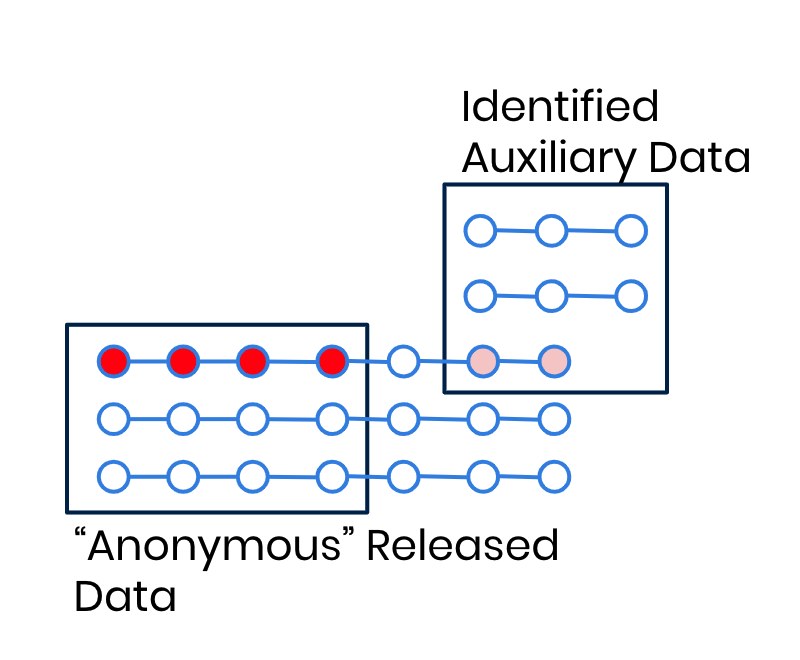} }}%
    \caption{Angle of Attack}%
    \label{fig:attack_types}%
\end{figure}

The above attack scenario requires that there is a temporal overlap (coinciding events) between the released and the auxiliary dataset. We take this further, and show that a successful linkage doesn't necessarily rely on shared records, but can also be performed by extracting and comparing behavioral patterns among disjoint set of records. Any existence of overlapping data points, particular at subject-level like basic sociodemographic attributes, would further increase the likelihood of success. The attack scenario (see Figure~\ref{fig:attack_types}) is outlined as follows:

\begin{enumerate}
  \item Organization releases an "anonymous" behavioral dataset D for period P1
  \item Attacker learns characteristic patterns of individuals from D
  \item Attacker obtains auxiliary data A of a known user X for period P2
  \item Attacker extracts and matches the characteristic patterns of D and A
  \item If successful, attacker can then reveal activities of user X within D
\end{enumerate}

\cite{vamosi2021} introduced an AI model, based on a Triplet-Loss Recurrent Neural Networks (TL-RNN), that can learn and abstract characteristic patterns of individuals from sequential data. That model optimizes for an embedding space, so that two different sequences of the same user end up closer together than two sequence of two different users. A similar approach has been introduced and shown to be effective for face recognition in \cite{schroff2015facenet}. And the same idea can now be applied to re-identify individuals based on behavioral categorical data, i.e., based on sequences of structured data records. Once the model is being trained, any sequence of data can then be mapped into a corresponding embedding space, yielding a numeric latent vector that represents the most characteristic traits of that sequence. The re-identification can then be simply performed via a Nearest-Neighbor Search in that embedding space.

\section{Empirical Re-Identification Study}

For our empirical demonstration, we use 2018 data from the Comscore Web Browser Panel, which provides continuous tracking of the visited websites among its panelists. We assume that only the visited sequence of domains (i.e., no timestamps, no visit duration, etc.) for January to June data (P1) is being released for 4'000 active, yet “anonymous” panelists, and we attempt to re-identify 1'000 individual panelists based on their observed July data (P2). Note, that there is no overlap in period P1 and P2, and each individual exhibits a different set of records during these periods. The question is then, whether the patterns within the data are characteristic and specific enough, so that we are still able to link individuals across periods. In total we have between 500 and 2'500 visits per panelist recorded, with an average of 1'290 visits, all across 115'000 distinct domains.

First of, the TL-RNN model is fitted to P1 data, with the records of each panelist being split by calendar week into several sequences. The Triplet Loss then optimizes the weights of the neural network, so that the week-wise records of the same panelist are embedded in its latent space closer together than the records of different panelists. The fitted TL-RNN model is then used to map the last week of P1, and the full July information of P2 into its 128 dimensional embedding space, to then search for nearest neighbors therein.

A pure random guess would result in only a 1 in a 4,000 chance of a successful match, i.e., a 0.025\% probability of identifying the correct P1 panelist for a given P2 panelist. However, after training the TL-RNN model on the P1 data of those 4'000 customers, and use the corresponding embeddings to identify nearest neighbors of P1 sequences and P2 sequences, we are able to correctly link 499 customers - thus have a success of re-identification of 49.9\%. Relaxing the success criteria by considering the five nearest neighbors within P1, then we find matches for 65.6\% of the panelists. This demonstrates that the re-identification on behavioral traits alone, without any overlap in data points, is indeed possible and becoming practical. The addition of any overlapping information, like subject-level attributes, would only further boost the rate for a successful match.

The next question is then, whether a data publisher can prevent re-identification by injecting random noise to the released dataset D. While this has already been shown to be of limited effect to protect against linkage attacks \cite{narayanan2006break}, a pattern attack is expected to be even more resilient against noise. We thus constructed a perturbated dataset D' by replacing any data point of an individual with 30\% probability with a data point from any other subject. Even though this is already a highly destructive mechanism, 26.6\% of individuals could still be re-identified. Even when going up to 60\% of data points being randomly permutated, 1.1\% of individuals were successfully re-identified. Thus, the presented type of re-identification attack is robust against noise, where all results of the re-identification task are presented in Table \ref{tab:nn}.

\renewcommand{\arraystretch}{1.5}
\begin{table}
\begin{center}
  \begin{tabular}{ p{2.5cm} p{2.3cm} p{2.3cm} p{2.3cm} p{2.3cm}}
  \hline \hline
    Permutation & N1N & N3N & N5N & N10N \\
  \hline
    0\% & 49.9\% & 60.6\% & 65.6\% & 71.5\% \\
    \hline
    10\% & 42.8\% & 52.0\% & 56.4\% & 63.8\% \\
    \hline
    20\% & 34.0\% & 46.5\% & 52.9\%  & 60.4\% \\
    \hline
    30\% & 26.6\% & 36.4\%  & 40.5\% & 48.7\% \\
    \hline
    60\% & 1.1\% & 1.4\%  & 1.5\% & 1.8\% \\
   \hline \hline
  \end{tabular}   
\caption{Ten, five, three and nearest neighbor(s) that include the correctly re-identified individual. In percent of totally 1,000 auxiliary users.}
\label{tab:nn}
\end{center}
\end{table}

A completely different approach towards privacy-safe sharing of granular-level data is the emerging domain of AI-based data synthetization. A generative model is being fitted to the original dataset, that then allows an arbitrary number of new, yet statistically representative data records to be created. Synthetic data is particularly known for its applications for image \citep{karras2017progressive} and text \citep{brown2020language} generation, but can be applied to structured behavioral data \citep{lin2020using} just as well. As there is no 1:1 relationship between actual and synthetic subjects, the re-identification is by definition not possible. Yet, the learning algorithm might still leak information on individuals into the released dataset via memorization, and thus empirical validation methods to assess the privacy of synthetic data are being introduced (see \cite{platzer2021holdout} and \cite{alaa2021faithful}). For this study, we pick up on the privacy concept, that the synthetic subjects shall not be any closer to the training subjects than to holdout subjects. And the embedding space of TL-RNN provides us with the relevant, task-specific distance measure, yielding a sensitive metric towards subject-level memorization.

\renewcommand{\arraystretch}{1.5}
\begin{table}
\begin{center}
  \begin{tabular}{p{1cm} p{0.82cm} p{0.82cm} p{0.82cm} p{0.82cm} p{0.82cm} p{0.82cm} p{0.82cm} p{0.82cm} p{0.82cm} p{0.82cm}}
  \hline \hline
    Flip & 10\% &20\%&30\%&40\%&50\%&60\%&70\%&80\%&90\%&synt.\\
  \hline
    DCR share & 98.5\% &89.9\%&70.6\%&62.6\%&58.7\%&58.1\%&57.6\%&57.5\%&55.4\%&51.8\% \\
   \hline \hline
  \end{tabular}   
\caption{Privacy Test based on Embedding}\label{dcr_share}
\label{tab:userclust}
\end{center}
\end{table}

Thus, we split the June data of the 4,000 panelists into equally sized training and holdout set, and generate 2,000 synthetic subjects based on the former with a publicly available synthetization software\footnote{https://mostly.ai/}. We then measure the distances of each synthetic subject to its nearest training subject, as well as to its nearest holdout subject within the previously constructed TL-RNN embedding space. The resulting average distance to training is 0.731, which is near identical to the average distance to holdout of 0.737. Another metric to investigate, is the share of synthetic records that ends up closer to a training subject than to a holdout subject. As we've split the original dataset 50/50, but are not using the holdout for synthetization, the expected lower bound for that share is at 50\%. In our empirical study, 51.8\% of synthetic subjects end up closer to a training subject, and resp. 48.2\% closer to a holdout. Thus, the synthetic records are shown to be nearly just as likely close to a training subject than close to a holdout subject, that was never seen before.

\begin{figure}[!ht]
\begin{center}
\includegraphics[width=14cm]{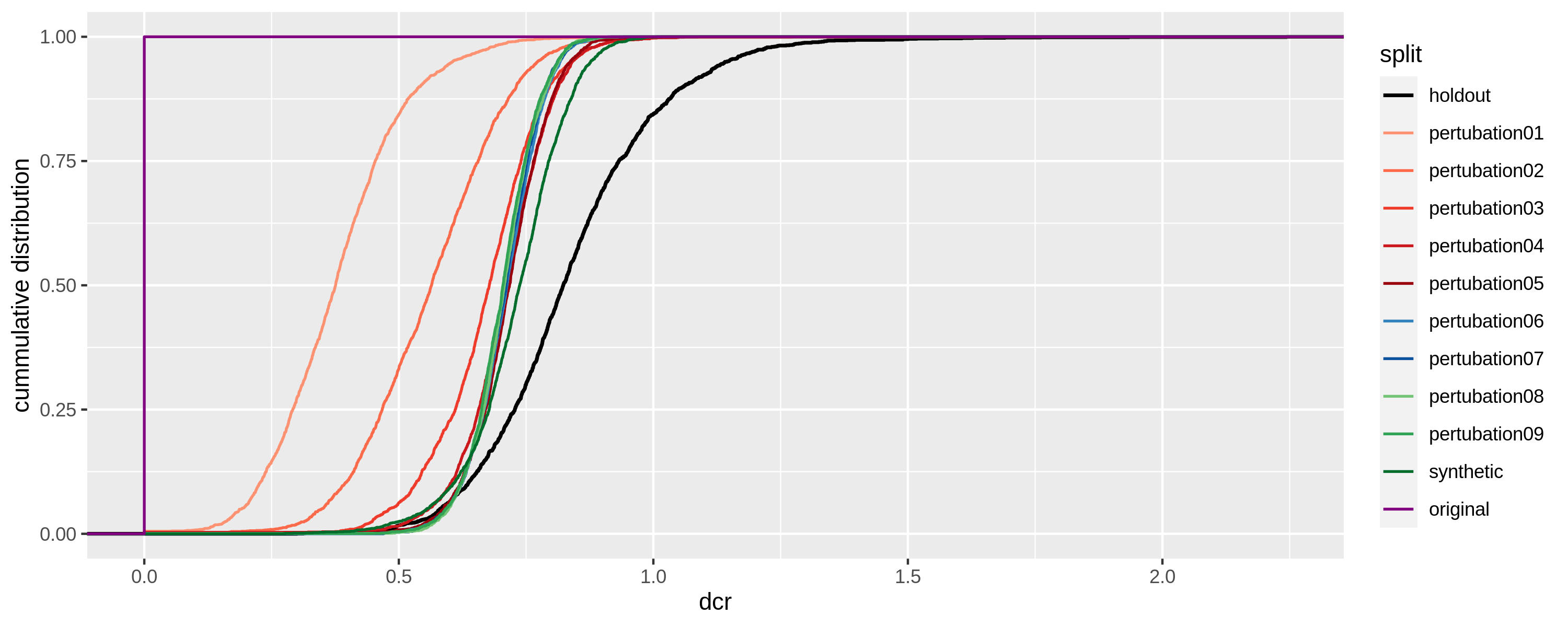}
\end{center}
\caption{Cumulative distribution function of the distance to closest ratio for different permutations, synthetic data and holdout.}\label{dcr_dist}
\end{figure}

This last metric is particularly remarkable, as even for a perturbated dataset, for which 90\% of data points were being randomly replaced, we still have 57.5\% of records being closer to the original dataset, than to another records. According to Figure \ref{dcr_share}, its cumulative distribution function confirms the conclusion, that synthetic data set is the closest to the holdout compared to perturbated records.

\section{Conclusions}

We have demonstrated how novel AI models can be leveraged to successfully re-identify behavioral data. In addition, we have shown that basic anonymization techniques, such as random perturbation, offer little safety against such pattern attacks. With these methods becoming increasingly stronger, and more commonly available, this has wider implication for what data can still be considered truly anonymous in the context of modern-day privacy regulations. Finally, we have also shown that data synthetization can offer a viable alternative to releasing data, as it defies even strong empirical privacy evaluations.

\bibliography{iclr2021_conference}

\end{document}